\documentclass[twocolumn,prl,superscriptaddress,floatfix,aps]{revtex4-1}

\usepackage{amsmath}
\usepackage{amssymb}
\usepackage{tabularx}
\usepackage{makecell}
\usepackage[dvipsnames]{xcolor}
\usepackage{graphicx}
\usepackage{bm}
\usepackage{siunitx}
\usepackage{hyperref}
\usepackage{chemformula}
\usepackage{mdframed}
\usepackage{listings}
\usepackage{verbatim}

\usepackage{xcolor}
\hypersetup{
    colorlinks,
    linkcolor={red!50!black},
    citecolor={blue!50!black},
    urlcolor={blue!80!black}
}

\date{September 2022}

\begin{document}
\title{
Vector graphics extraction and analysis of electrical resistance data\\
in Nature volume 586, pages 373–377 (2020)
}
\author{J.\ J.\ Hamlin}
\affiliation{Department of Physics, University of Florida, Gainesville, FL 32611, USA}

\begin{abstract}
In this paper, I present an analysis of the electrical resistance graphs in Ref.~\cite{Snider2020}, which reported the discovery of room temperature superconductivity in a carbonaceous sulfur hydride and was subsequently retracted on September 26th, 2022.
I show that, over a single temperature interval, the electrical resistance data can be decomposed into at least two signals of differing digital precision, thus raising questions concerning the methods used to obtain the published data.
Since the raw data-files for the electrical resistance measurements have not been made available, in order to perform this analysis, I have developed a set of python scripts to extract the data-points with high precision from the internal structure of the vector graphics image files.
I describe the data extraction method.
Example code and the resulting electrical resistance vs temperature data-files are made available in public repositories.
\end{abstract}

\maketitle

\section{Introduction}
In October 2020, a paper was published reporting the discovery of room temperature superconductivity in a carbonaceous sulfur hydride (CSH) under high pressure conditions~\cite{Snider2020}.
The claim was based primarily on three  pieces of evidence: diamagnetic transitions in the ac magnetic susceptibility, zero electrical resistance, and suppression of the transition in an applied magnetic field.
On Monday, September 26th, a retraction was published, with the notice focusing on the validity of the background subtraction method applied to the ac magnetic susceptibility data.
The retraction notice notes that the authors maintain that the raw data provide strong support for the main claims of the original paper.

In this work, I have analyzed the electrical resistance graphs in Fig.~1a and Fig.~2b of Ref.~\cite{Snider2020}, which report electrical resistance measurements.
Hereafter, I refer to these figures as CSH-Fig1a and CSH-Fig2b, respectively.
I find artifacts similar to those that appear in the magnetic susceptibility data~\cite{vanderMarel2022}.
In the case of the magnetic susceptibility data, two of the authors of Ref.~\cite{Snider2020} stated that such artifacts are a consequence of the user-defined background that was subtracted from the data~\cite{reply}.
No such background subtraction for the published electrical resistance data was indicated in Ref.~\cite{Snider2020}.

\section{Vector-based data extraction}
The authors of Ref.~\cite{Snider2020} have not made the electrical resistance raw data files publicly available.
In such situations interested parties are left to analyze the published graphs.
Typically, this is done using bitmap-based data extraction methods implemented by programs such as DataThief~\cite{DataThief} and Webplotdigitizer~\cite{Rohatgi2022}.
In this method, the axes scale bars are manually calibrated and the data points are selected either by hand or, in some cases, using automated methods that function with varying levels of success.
Bitmap-based data extraction faces several limitations.
The precision with which the data points can be extracted is limited by the resolution of the image file.
In favorable cases, the image resolution may be approximately 600 pixels-per-inch.
However, figures are often included in formats such as \verb!jpg!, where image compression may introduce artifacts that blur the edges of data points.
Most importantly, data points that are visually covered by other data points will be invisible and thus impossible to extract.
This is often the case for graphs that contain a large number of densely-spaced data points.

Many journals adopt policies that graphics should be included in vector format when possible.
Common vector graphics formats include \verb!pdf!, \verb!eps!, and \verb!svg! (scalable vector graphics).
A simple way to determine whether a graph in a paper is a vector-based image is to use the text-selection tool to try to select some of the text in the graph.
If you are able to select the text, it is likely a vector-based image.
A vector format image will remain sharp no matter how much you zoom-in on it.
Rather than describing graphics as a grid of pixels, these formats can include instructions describing a set of paths to draw.
The full \verb!svg! instruction specification is available at Ref.~\cite{svg}.
For example, a square data point might be described by specifying the coordinates of a set of four lines.
Figure~\ref{fig:svg_code} shows examples of the path instructions for (a) a single red, Run 2 data point in CSH-Fig1a and (b) a single red \SI{9}{T} data point in CSH-Fig2b, after converting the corresponding pages of the article \verb!pdf! into \verb!svg! format using \verb!pdf2svg!~\cite{pdf2svg}.
Typically, the coordinates associated with these paths are stored with substantially higher precision than a corresponding bitmap equivalent.
\begin{figure*}
    \centering
    \begin{mdframed}
    \vspace{1em}
    \textbf{(a)}
    {\footnotesize
        \begin{lstlisting}
<path style=" stroke:none;fill-rule:evenodd;fill:rgb(94.117737%,32.156372%,32.548523%);fill-opacity:1;" d="M 208.292969 217.472656 C 208.292969 218.171875 207.726562 218.738281 207.03125 218.738281 C 206.332031 218.738281 205.765625 218.171875 205.765625 217.472656 C 205.765625 216.777344 206.332031 216.210938 207.03125 216.210938 C 207.726562 216.210938 208.292969 216.777344 208.292969 217.472656 "/>
        \end{lstlisting}
    }
    \hrule
    \vspace{1em}
    \textbf{(b)}
    {\footnotesize
        \begin{lstlisting}
<path style="fill-rule:evenodd;fill:rgb(92.941284%,13.33313%,14.117432%);fill-opacity:1;stroke-width:0.139;stroke-linecap:round;stroke-linejoin:round;stroke:rgb(52.941895%,9.01947%,9.803772%);stroke-opacity:1;stroke-miterlimit:10;" d="M 0.00045777 -0.00143164 L 1.340302 1.338412 L 2.684052 -0.00143164 L 1.340302 -1.337369 Z M 0.00045777 -0.00143164 " transform="matrix(1,0,0,-1,463.726105,81.529818)"/>
        \end{lstlisting}
    }
    \end{mdframed}
    \caption{Example of instructions used to draw data points in SVG files.
    (a) Shows an example of a red, Run 2 data point from CSH-Fig1a, while
    (b) shows an example of red \SI{9}{T} data point from CSH-Fig2b.
    Note the different structure of the two sets of instructions: (b) uses a transform matrix, while (a) does not.
    Different graphs may describe the data in slightly different ways, but the instructions can be straightforwardly parsed to extract the coordinates of each data point.}
    \label{fig:svg_code}
\end{figure*}

In principle then, it is possible to analyze the instructions in a vector image file to extract the coordinates of the data points.
Extraction of the data points from the vector image data allows extraction of points that are completely covered by other data points and are therefore not directly visible when viewing the image in a \verb!pdf! of a manuscript.
Such is not possible from a bitmap image.
While not common, vector-based data extraction has been implemented and shown to be reliable in least one other study~\cite{extraction}.
The technique has the potential to aid in future data-mining efforts.
Databases of experimental information such as the The Pauling File~\cite{Villars2019}, may consider the feasibility of adding vector-extracted data to their collections.

The data extraction workflow employed in this work is as follows:
\begin{enumerate}
 \item Convert an entire page of the \verb!pdf! into \verb!svg! format using the \verb!pdf2svg! utility~\cite{pdf2svg}.
 \item Identify the hexadecimal color for each data set.
 \item Use the svgpathtools python library~\cite{svgpathtools} to extract the \verb!svg! path objects that have the correct color and compute the centroid of the vertices of each path (i.e.\ the center of each data point), or in the case of line plots, the coordinates of each line segment.
 \item Remove data points from the series that come from the figure legends or other figures on the same page.
 \item Analyze the locations of the tick marks in the \verb!svg! file in order to calibrate the scale.
\end{enumerate}
Full details of the data extraction method and example code are available in extensively commented jupyterlab notebooks hosted on github~\cite{github} and archived on zenodo~\cite{zenodo}.
The program Inkscape~\cite{Inkscape} can also be useful for initial exploration of the structure of the \verb!svg! file.
Inkscape's \verb!pdf! to \verb!svg! conversion assigns a path identification (ID) number property to every path, which can be helpful for identifying the parts of the \verb!svg! code that correspond to certain objects in the graph.
One can open the \verb!svg! in Inkscape, click on an item to find the path ID, and then open the \verb!svg! in a text editor and search for that path ID to view the corresponding code.

\section{Validation}
In order to validate the extraction method, I include here an analysis of the extracted data for Fig.~3 of Ref.~\cite{Lim2021} (Re22Be-Fig3), for which I have access to the original raw data files.
The paper was recently published in Physical Review B, and Re22Be-Fig3 was included as a vector graphics image.
This figure presents normalized electrical resistivity vs temperature measured at several pressures for the material \ch{Be22Re}.
Superconducting transitions are present in each measurement.
For this analysis, we focus on the measurement at \SI{15}{GPa}, because at that pressure, the data at both high and low temperature are mostly obscured by measurements at other pressures.
Bitmap-based data extraction would not be able to recover data in those regions.

\begin{figure}
    \centering
    \includegraphics[width=\columnwidth]{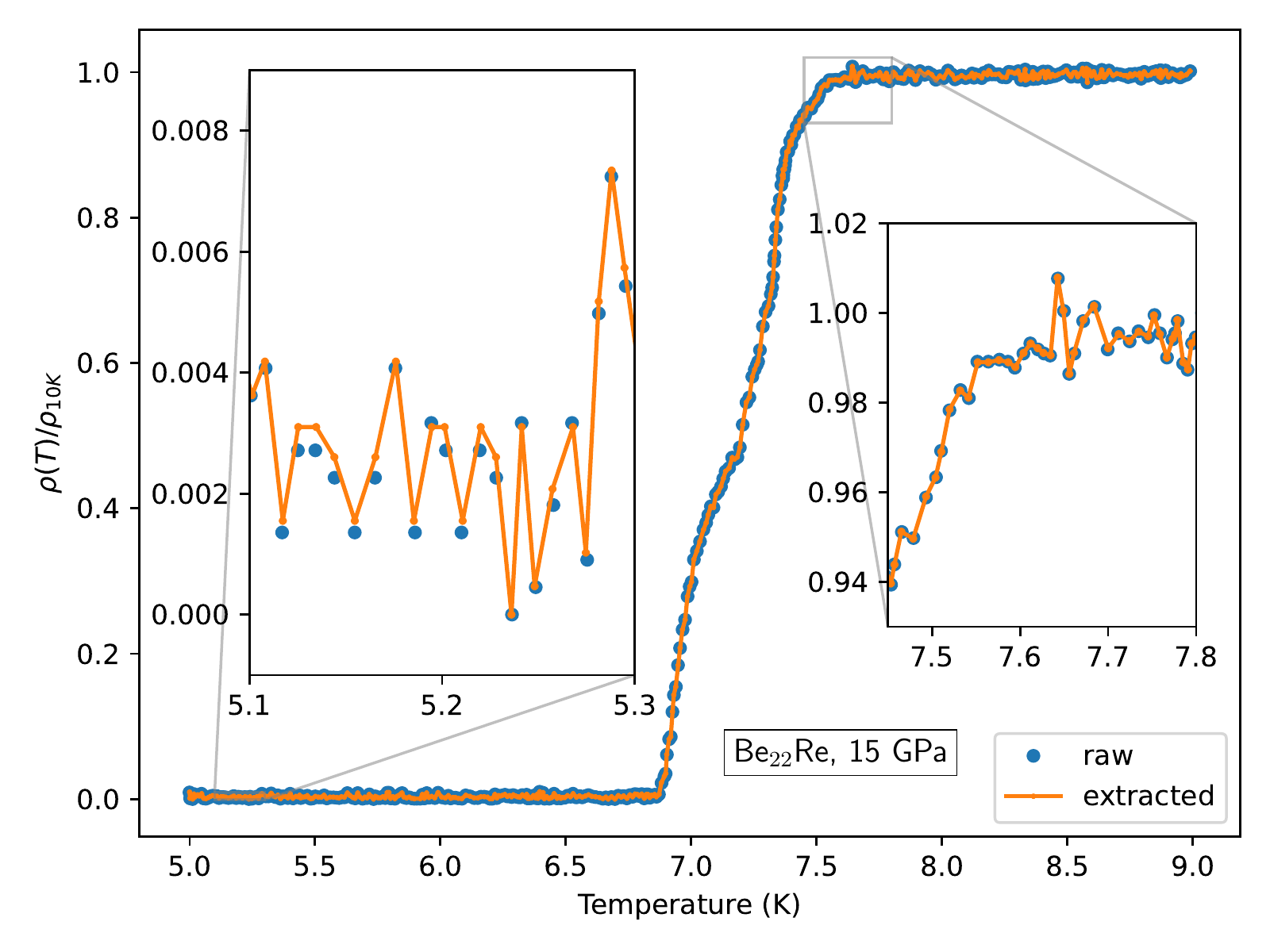}
    \caption{Comparison between data extracted from Be22Re-Fig3 (Fig.~3 of Ref.~\cite{Lim2021}) and the original, source raw data.
    Data shown is for \ch{Be22Re} at a pressure of 15 GPa.
    The raw resistance data have been normalized to the value at 10 K.
    In both regions highlighted in the insets, the \SI{15}{GPa} data is largely obscured by measurements at other pressures in Be22Re-Fig3.
    Nevertheless, the data can be accurately recovered in the obscured regions.}
    \label{fig:Be22Re}
\end{figure}

Figure~\ref{fig:Be22Re} shows a comparison of the extracted data and the original raw data.
The raw resistance data has been normalized to the value at \SI{10}{K}, consistent with the axis labeling in Be22Re-Fig3.
The insets of Fig.~\ref{fig:Be22Re} show zoomed comparisons of the extracted and raw data.
Very small differences between the two data sets are caused by the limited precision of the \verb!svg! path data.

\section{Precision of extraction}
While a vector image plot in a publication has higher precision than a corresponding bitmap, the precision is not unlimited.
The precision of data in a raw data file is set not only by the number of digits recorded in the file but also by the number of bits in the analog to digital converter used by the instrument.
For extracted data, precision is determined both by the precision of the underlying raw data and also by the number of digits used to store the coordinates of the various paths.
In all cases, floating point errors may also be present~\cite{Goldberg1991}.

For the normalized resistivity, the raw data has a precision corresponding to \num{4.5e-4}, while the extracted data exhibits a coarser precision of \num{5.3e-4}.
For the temperature values, the raw data has a precision corresponding to roughly \SI{7.3e-4}{\kelvin}, while the extracted data exhibits a coarser precision of \SI{1.3e-3}{\kelvin}.
The precision of the raw temperature values is indicated as a ``rough'' value, because what is actually measured is the resistance of a calibrated Cernox thermometer.
The ratio between the (semiconducting)
Cernox thermometer resistance and the corresponding temperature is non-linear.
Examination of the left inset of Fig.~\ref{fig:Be22Re} shows that the agreement between raw and extracted data is fully consistent with the values of precision stated above.
The precision of the extraction is equivalent to a resolution of $\sim 1000$ pixels-per-inch, substantially higher than available for most figures included in papers as \verb!jpg! images.

\section{Extraction of CSH data}
\begin{figure}
 \centering
 \includegraphics[width=\columnwidth]{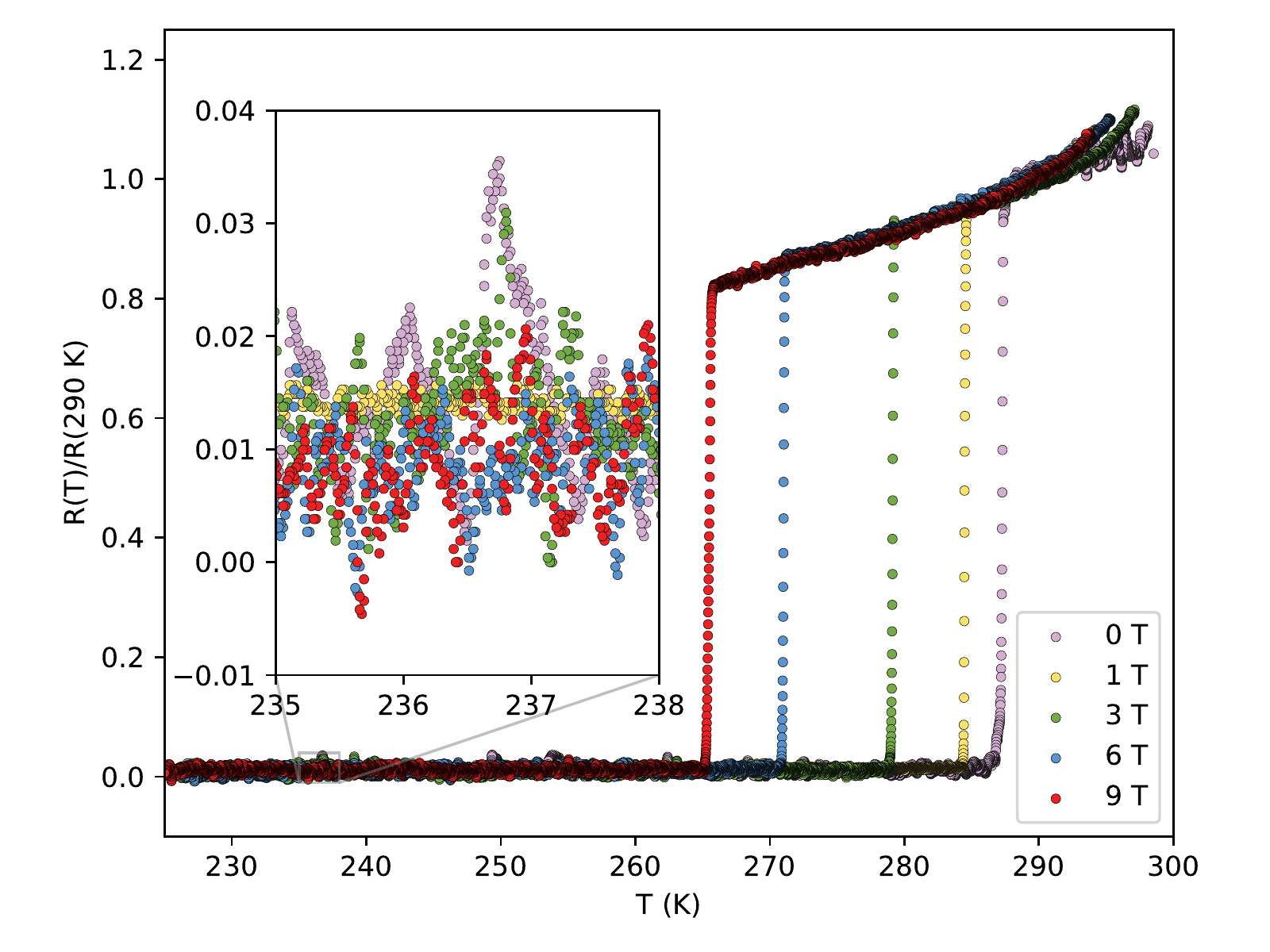}
 \caption{The data extracted from CSH-Fig2b (Fig.\ 2b of Ref.~\cite{Snider2020}), which show field dependent resistivity at \SI{267}{GPa}.
 The zoomed region shows that, in contrast to bitmap-based extraction, our method allows overlapping points to be extracted.}
 \label{fig:Fig2b_replica}
\end{figure}

Using the methods described above, I extract roughly 40,000 data points from CSH-Fig1a and 30,000 points from CSH-Fig2b.
These figures correspond to electrical resistance vs temperature at different pressures and at different applied fields, respectively.
Figure~\ref{fig:Fig2b_replica} shows the data extracted from CSH-Fig2b.
Readers may compare Fig.~\ref{fig:Fig2b_replica} with the original CSH-Fig2b.
Viewing both figures at a high level of zoom, it is apparent that all the subtle features of the noise are reproduced.
Files containing the extracted data points are published on Zenodo~\cite{zenodo}.

\begin{figure}
 \centering
 \includegraphics[width=\columnwidth]{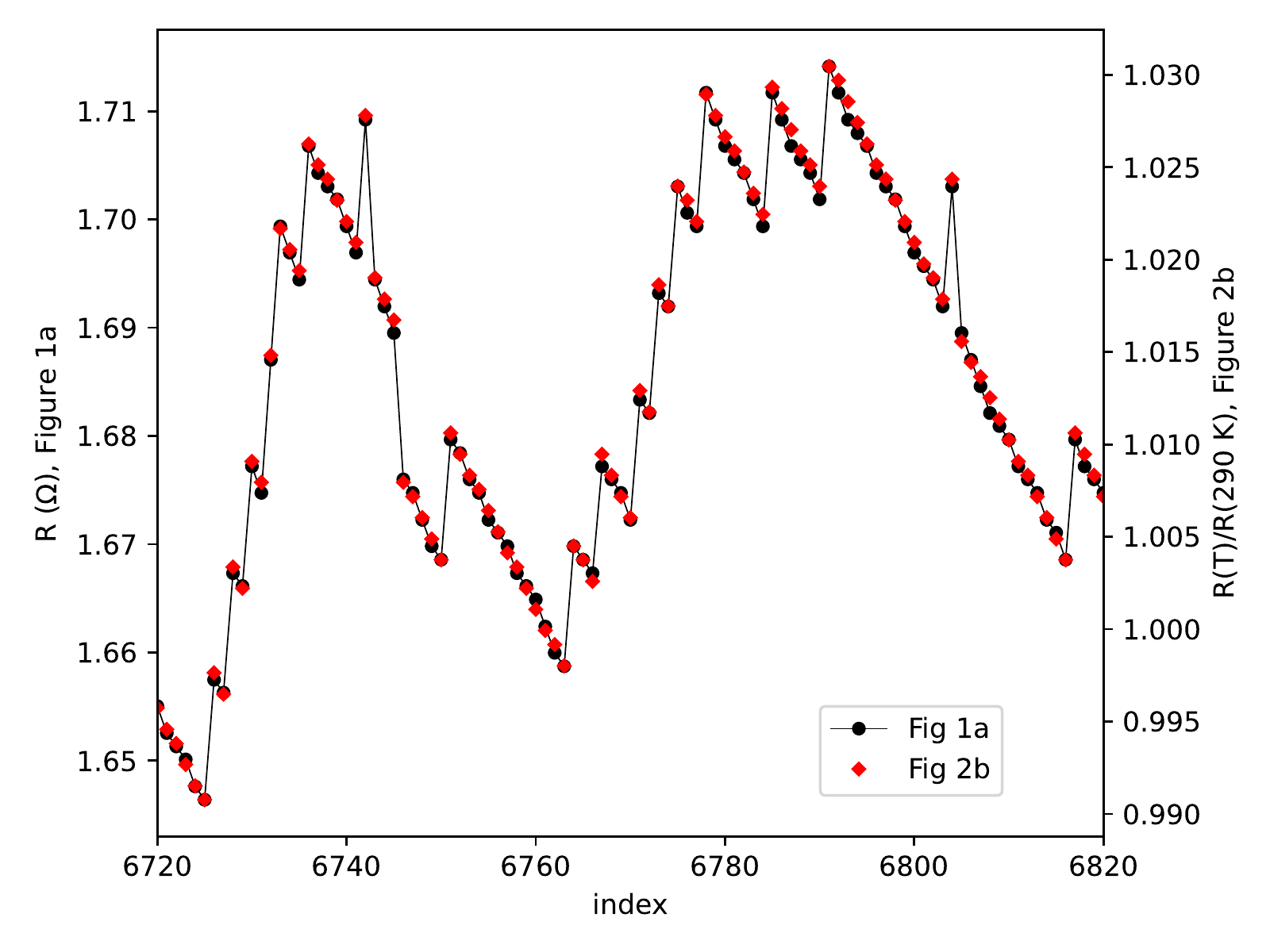}
 \caption{
 Comparison of 267 GPa data extracted from CSH-Fig1a and CSH-Fig2b.
 To highlight the very good agreement, shown here are data over a very narrow temperature region (291.7 K to 291.8 K).
 The data show an unusual structure of discrete jumps that are intermixed with more smoothly varying data.
 Typically, raw data from a digital instrument exhibit discrete jumps with \textit{constant} valued data in between the jumps.
 }
 \label{fig:extraction_comparison}
\end{figure}

In order to gain confidence that the data have been reliably reconstructed, we can compare data extracted from CSH-Fig1a and CSH-Fig2b.
Both of these figures present the same 267 GPa data set, upon which the claim of room-temperature superconductivity was based.
Figure~\ref{fig:extraction_comparison} shows the 267 GPa data extracted from each of these two figures over a narrow temperature range corresponding 291.7 K to 291.8 K. 
Note that the scales of the left and right y-axis have been adjusted to align the two data sets, since they were reported in different units (ohms vs normalized resistance).
The fidelity of the extraction is clearly good, as the two data sets agree closely.

In the case of CSH-Fig2b, the values of temperature are rounded to increments of $\sim \SI{17}{mK}$ while the values of $R(T)/R(\SI{290}{K})$ are rounded to increments of 0.0004 (roughly the symbol size in Fig.~\ref{fig:extraction_comparison}).
Note that the precision of the temperature values extracted from CSH-Fig1a are significantly lower since that plot covers a much larger range of temperatures in the same physical dimensions on the page.

\section{Analysis extracted CSH data}
All raw data that have been collected using computers and digital equipment (in contrast to analog instrumentation connected to chart recorders) will present some degree of digitization.
When changes in the underlying measured signal are small compared to the instrumental precision, the collected data will present as a series of constant-valued data points, interrupted by discrete jumps of constant magnitude.
The CSH data shown in Fig.~\ref{fig:extraction_comparison} instead show a smoothly varying signal interrupted by discrete jumps.
It appears that over a single temperature interval the signal can be described by at least two components of differing digital precision.

\begin{figure}[htp]
 \centering
 \includegraphics[width=\columnwidth]{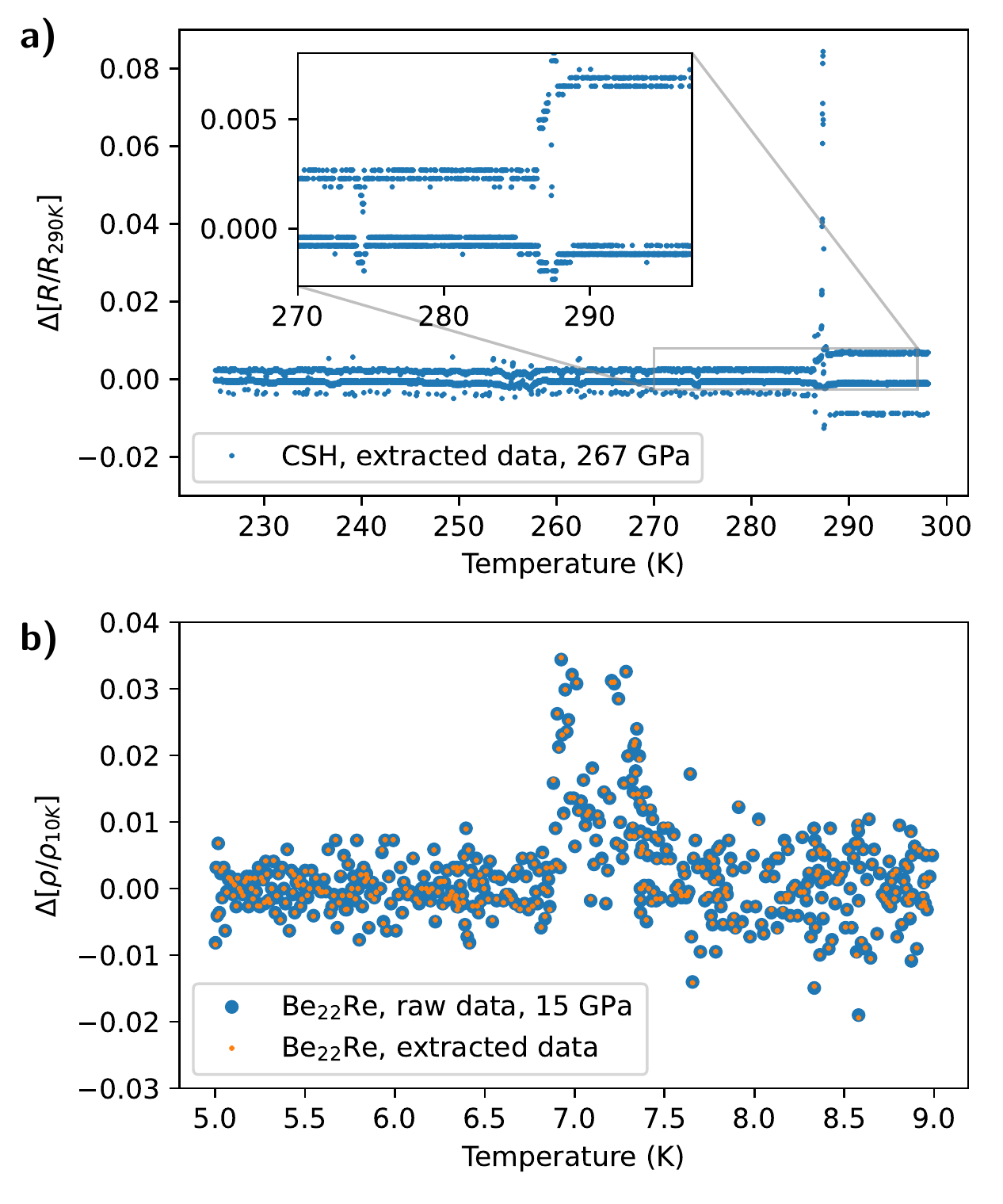}
 \caption{Difference between adjacent data points vs temperature for a) the 267 GPa data from CSH-Fig2b~\cite{Snider2020} and b) the 15 GPa data from Be22Re-Fig3~\cite{Lim2021}.
 The CSH normalized resistance data shows digitization at the underlying precision of the extraction process (0.0004) but also at two other increments, 0.0078 and 0.0031 at high and low temperatures, respectively.
 }
 \label{fig:diff}
\end{figure}

The pattern of smooth segments broken by discrete jumps for CSH-Fig2b can be visualized by looking at a ``difference'' plot of $\Delta (R/R_{290 K})$ where
\begin{equation}
 [\Delta (R/R_{290 K})]_i = [R/R_{290 K}]_i - [R/R_{290 K}]_{i-1}.
\end{equation}
This quantity is plotted in Fig.~\ref{fig:diff}a.
The narrow peak near 287 K is the superconducting transition.  At high temperature, the data show jumps of $\pm 0.0078$ (see inset).
Note that this is more than an order of magnitude larger than the precision of the vector extraction (0.0004). The level of digitization in the portion of the data below $T_c$ is different and equal to $\sim \pm 0.0031$.
The appearance of Fig.~\ref{fig:diff} derives from the structure of the data shown in Fig.~\ref{fig:extraction_comparison}.
For example, at higher temperatures, we see three roughly flat lines.
The middle line corresponds to the more smoothly varying portions of the normalized resistance in between the discrete jumps.
The upper and lower lines correspond to the places where the normalized resistance jumps up or down by the increment of 0.0078.
For comparison, Fig.~\ref{fig:diff}b shows difference data for the \SI{15}{GPa} data from Be22Re-Fig3, which shows none of the unusual artifacts that are present in panel a).

\section{Unwrapping CSH data}
Van der Marel and Hirsch~\cite{vanderMarel2022} noted a similar structure to the susceptibility data of Ref.~\cite{Snider2020}, and described an ``unwrapping'' procedure to disentangle the smoothly varying and digitized parts of the susceptibility data.
Here, I apply a similar method to the resistance data.
I have implemented an unwrapping algorithm~\cite{github} that is straightforward and involves shifting the resistance values (or in this case $R/R_{290 K}$ values) up or down by integer multiples of the digitization increment in order to minimize the changes in $\Delta(R/R_{290 K})$.

I perform the unwrapping procedure using a digitization increment of 0.0078 for the data above $T_c$.
In the region close to $T_c$ it is difficult to identify which step size should be used and at low temperatures the digitization is close enough to the precision of the underlying vector graphics path data that minor ambiguities are introduced into the unwrapping procedure.
Therefore, I have left out discussion of the unwrapped data below 288 K from this report.
It may be possible to perform a similar analysis on the portion of the data below $T_c$ using the source data, if they were made available.
The source data should have a higher precision than the extracted data.

\begin{figure}
 \centering
 \includegraphics[width=\columnwidth]{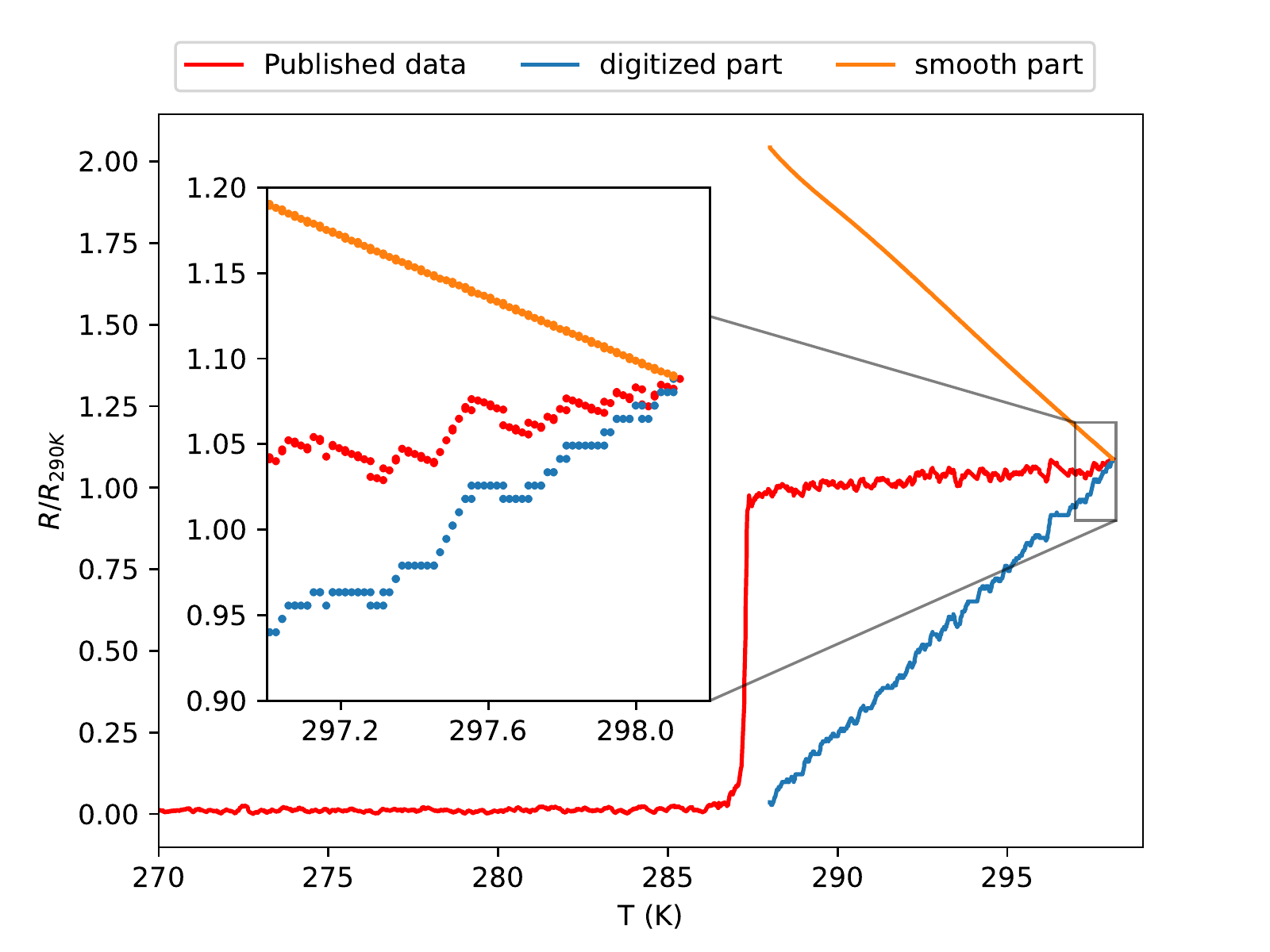}
 \caption{The 267 GPa data, decomposed (unwrapped) into digitized and smooth components.
 The blue, digitized component resembles typical raw data collected from a digital instrument (inset).
 }
 \label{fig:uf_high}
\end{figure}

\begin{figure}
 \centering
 \includegraphics[width=\columnwidth]{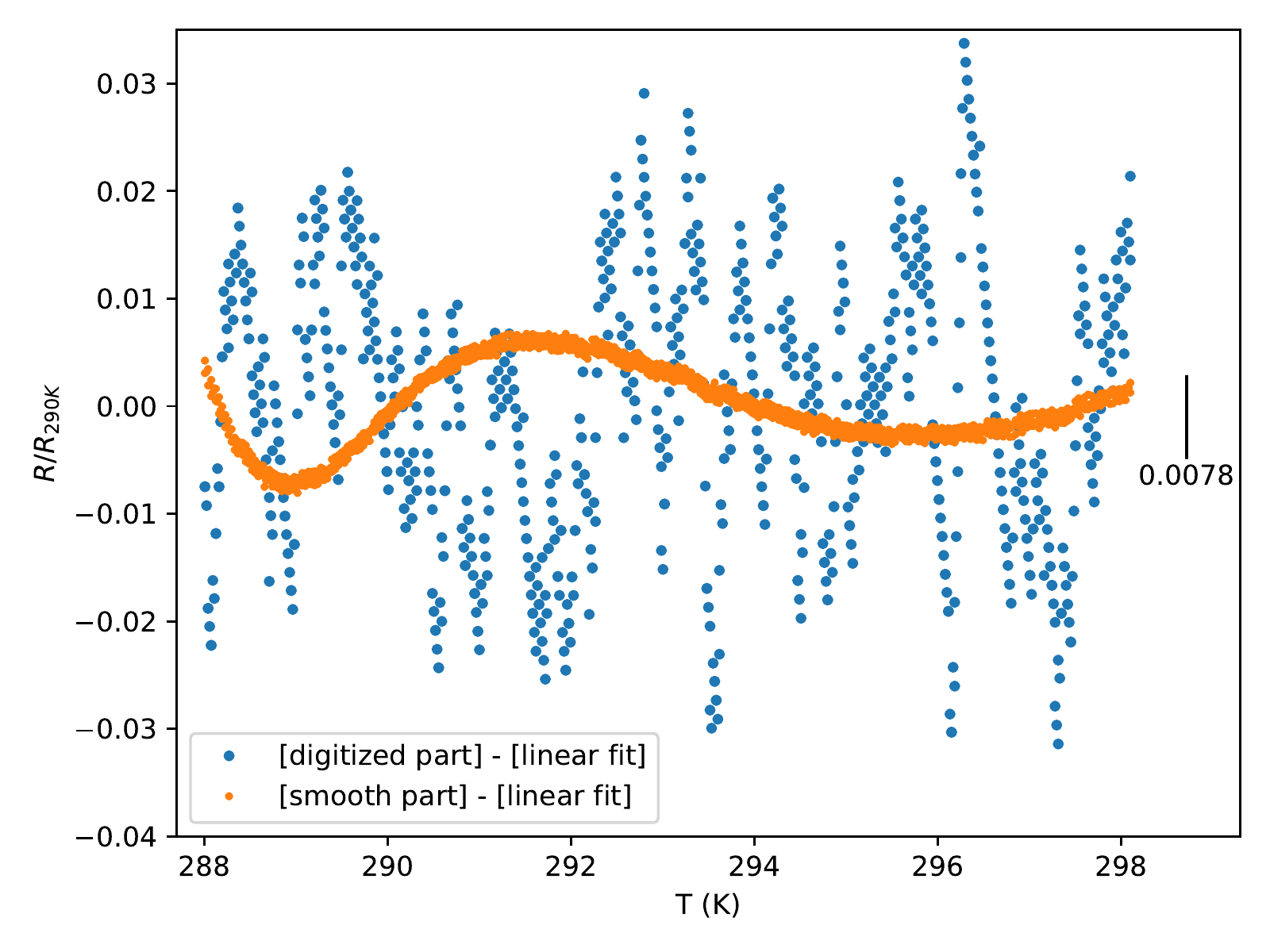}
 \caption{Comparing the digitized component with the smooth component of the published data above $T_c$, where a linear fit has been subtracted from each data set in order to highlight the deviations from linearity.
 }
 \label{fig:linsub_high}
\end{figure}

Figure~\ref{fig:uf_high} shows the result of unwrapping the portion of the data above $T_c$.
The published data can be arrived at by adding the orange curve to the blue digitized component.
The change in digitized component from room temperature down to $T_c$ is large compared to the full-scale range of the published data.
The orange, smoother component of the data is presented in greater detail in Figure~\ref{fig:linsub_high}, 
where a linear fit has been subtracted in order to highlight the deviations from linearity.
Note that the scatter in the smooth component is significantly smaller than the increment of digitization present in the published data (0.0078).
Unwrapping the data with an increment different than 0.0078 typically produces a noisy, random curve.
Note that the smooth component exhibits some scatter, of order four times the precision  (0.0004) of the extracted data.
A similar analysis beginning from the original source data files that formed the basis of CSH-Fig2b would allow a more detailed analysis of this scatter.

\section{Conclusions}
I report a method for extracting data from published graphs, using the internal structure of the vector graphics instructions.
The method allows data to be extracted with significantly higher precision than typical bitmap-based methods.
Furthermore, data points that are obscured by overlapping data can also be recovered.
I apply the method to extracting electrical resistance data from Ref.~\cite{Snider2020}.
Analysis of the recovered data shows an unusual pattern of comparatively smooth data connected by discrete jumps, similar to what was found~\cite{vanderMarel2022} for the corresponding magnetic susceptibility data.
Some of the authors of Ref.~\cite{Snider2020} attributed~\cite{reply} such artifacts to the ``user-defined'' background subtraction method they applied to their susceptibility data,
However, no background subtraction was disclosed for the electrical resistance data.

These finding raise questions about the methods used to produce the resistance data that was published in support of the claim of room temperature superconductivity in CSH~\cite{Snider2020}.
The questions might be resolved by the release of the underlying raw data.
I have performed the same analysis on electrical resistance graphs from other publications on high pressure hydride superconductors and have \textit{not} observed similar artifacts.
The results highlight that, in the interest of data preservation, journals should require graphs to be submitted in vector rather than bitmap (raster) formats whenever possible.

\section{Acknowledgement}
This work was supported by the U.S.\ Department of Energy Basic Energy Sciences under Contract No.\ DE-SC-0020385.

\bibliographystyle{hieeetr}
\bibliography{references}

\end{document}